\documentstyle[amssymb,epsf,cite,11pt]{article}

\newcommand{\be}{\begin{equation}}
\newcommand{\ee}{\end{equation}}

\begin{document}

\title{\bf Multifractal analyses of daily rainfall time series in Pearl River basin of China}

\author{Zu-Guo Yu$^{1,2}$, Yee Leung$^3$\thanks{
  Corresponding author, email: yeeleung@cuhk.edu.hk}, Yongqin David Chen$^3$, Qiang Zhang$^4$, Vo Anh$^2$ ~and Yu Zhou$^{1}$\\
{\small$^1$ Hunan Key Laboratory for Computation and Simulation in
Science and Engineering and }\\
{\small Key Laboratory of Intelligent Computing and Information
Processing of Ministry of Education,}\\
{\small Xiangtan University, Xiangtan,  Hunan 411105, China.}\\
{\small $^{2}$School of Mathematical Sciences, Queensland University of Technology,}\\
{\small GPO Box 2434, Brisbane, Q4001, Australia.}\\
{\small $^3$Department of Geography and Resource Management, and
Institute of Environment, }\\
{\small Energy and Sustainability, The Chinese University of Hong
Kong, Hong Kong, China.}\\
{\small $^4$Department of Water Resources and Environment, and Key
Laboratory of }\\
{\small Water Cycle and Water Security in Southern China of
Guangdong High Education Institute,}\\
{\small  Sun Yat-sen University, Guangzhou 510275, China.}
 }
\date{}
\maketitle

\begin{abstract}
The multifractal properties of daily rainfall time series at the
stations in Pearl River basin of China over periods of up to 45
years  are examined using the universal multifractal approach
based on the multiplicative cascade model and the multifractal
detrended fluctuation analysis (MF-DFA). The results from these
two kinds of multifractal analyses show that the daily rainfall
time series in this basin have multifractal behavior in two
different time scale ranges. It is found that the empirical
multifractal moment function $K(q)$ of the daily rainfall time
series can be fitted very well by the universal mulitifractal
model (UMM). The estimated values of the conservation parameter
$H$ from UMM for these daily rainfall data are close to zero
indicating that they correspond to conserved fields. After
removing the seasonal trend in the rainfall data, the estimated
values of the exponent $h(2)$ from MF-DFA indicate that the daily
rainfall time series in Pearl River basin exhibit no long-term
correlations. It is also found that $K(2)$ and elevation series
are negatively correlated. It shows a relationship between
topography and rainfall variability.
\end{abstract}


\section{Introduction}

Rainfall is one of the most important variables studied because
its non-homogenous behavior in event and intensity, leading to
drought, water runoff and soil erosion with negative environmental
and social consequences \cite{Walther02, Valencia10}. Analysis and
modelling of rainfall are significant research problems in applied
hydro-meteorology \cite{Zhang08}. Rainfall time series often
exhibit strong variability in time and space.

Rainfall also exhibits scaling behavior in time and space (e.g.
[3-7]). There is thus a need to characterize and model rainfall
variability at a range of scales which goes beyond the scales that
can be directly resolved from observations \cite{Svensson96}.
Investigation of the existence of fractal behavior in rainfall
processes has been an active area of research for many years
\cite{Sivakumar00}. Some recent experiments have shown that scale
invariance, in time and space, does exist in rainfall fields
\cite{Michele05}. Olsson {\it et al}. \cite{Olsson93} investigated
the rainfall time series by calculating the box and correlation
dimensions via a monodimensional fractal approach (simple
scaling). Their results indicate scaling but with different
dimensions for different time aggregation periods. Hence the
investigated rainfall time series display a multidimensional
fractal behavior. Venugopal {\it et al}. \cite{Venugopal06}
employed the wavelet-based multifractal analysis to reexamine the
scaling structure of rainfall over time. Molnar and Burlando
\cite{Molnar08} used the exponent of correlation function, a
multifractal parameter, to study the seasonal and spatial
variabilities. Using 2-dimensional Fourier series analysis and
spectral analysis, Boni {\it et al.} \cite{Boni08} proposed a
methodology to study the estimated index factor for rainfall in
mountainous regions. During the past two decades, stochastic
models of rainfall have increasingly exploited the property of
multifractal scale invariance, resulting in multifractal models
that are more advantageous over conventional models in rainfall
representations [15-17].

   The multiplicative cascade model has been widely used to study the
multifractal properties of the rainfall data (e.g. [2, 4-8,
17-29]). Schertzer and Lovejoy \cite{Schertzer87} showed that
statistically scaled invariant processes are stable and converge
to some universal attractor, and thus can be defined by a small
number of relevant parameters, specifically three with the
universal multifractal framework.

The simple multifractal analysis (MFA) is based upon the standard
partition function multifractal formalism \cite{Halsey86},
developed for the multifractal characterization of normalized,
stationary measurements. Unfortunately, this standard formalism
does not give correct results for non-stationary time series that
are affected by trends or that cannot be normalized
\cite{Kantelhardt02}. Thus, two generalizations of simple MFA were
developed. One is the wavelet-based MFA which has been used to
study rainfall data (e.g. \cite{Venugopal06}). Another
generalization is the multifractal detrended fluctuation analysis
(MF-DFA) \cite{Kantelhardt02} which is an extension of the
standard detrended fluctuation analysis (DFA) introduced by Peng
{\it et al}. \cite{Peng92,Peng94}. DFA can be employed to detect
long-range correlations in stationary and noisy nonstationary time
series. It intends to avoid the unravelling of spurious
correlations in time series. The DFA method has been successfully
applied to problems in fields such as DNA and protein sequences
(e.g. \cite{Peng92, Yu01, Yu06}) and hydrology (e.g. [36-40]). The
MF-DFA is a modified version of DFA for the detection of
multifractal properties of time series. It renders a reliable
multifractal characterization of nonstationary time series
encountered in phenomena such as those in geophysics [31, 37, 38,
41-46]. The MF-DFA has also been successfully applied to problems
in hydrology (e.g. [37-39]). The relationship between topography
and rainfall variability is a very important issue in the study of
rainfall.

   Our work in this paper focuses on the multifractal properties of
daily rainfall time series and possible relationships between the
multifractal exponents and landscape properties. We use the
universal multifractal model (UMM) proposed by Schertzer and
Lovejoy \cite{Schertzer87} to fit the multifractal moment function
$K(q)$ of the rainfall data and propose a method to estimate the
parameters. We also adopt the MF-DFA approach to detect the
correlation and multifractal properties of daily rainfall data in
this paper.

As the largest watershed in South China, the Pearl River (Zhujiang
in Chinese) delta is a composite drainage basin with a total area
of 45.4$\times 10^4$ km$^2$, consisting of three major rivers
(i.e., West River, North River, and East River) and several
independent rivers in the downstream and delta regions (see Figure
1). The Asian monsoon and moisture transport are the important
influencing factors on precipitation patterns in this region.
Given its large size and dominance of a sub-tropical humid monsoon
climate, the Pearl River basin is under the influence of rainfall
variability which is a highly complicated process in space and
time. Zhang et al. \cite{Zhang2009a} reported an increased
high-intensity rainfall over the basin in conjunction with the
decreased rainy days and low-intensity rainfall. It was also found
that the abrupt changes of the precipitation totals (for annual,
winter, and summer precipitation) occurred in the late 1970s,
1980s, and early 1990s, and the precipitation intensity basically
increased after the change points \cite{Zhang2009a, Gemmer2011}.
In this paper, we study the daily rainfall data over the period
from 1 January 1960 to 31 December 2005 at 41 locations in Pearl
River basin using the UMM and MF-DFA methods. Parameters from the
above MFAs are used to infer the spatial relationship of rainfall
in Pearl River basin of China.

\section{ Multifractal analyses}
\subsection{Universal multifractal approach based on the multiplicative cascade model}

Let $T(t)$ be a positive stationary stochastic process at a
bounded interval of ${\bf R}$, assumed to be the unit interval (0,
1) for simplicity, with $E(T(t))=1$  (For a time series $x_i$,
$i=1,\cdots,L$, we can define $t_i=i/L$, and
$T(t_i)=x_i/(\sum_{k=1}^L x_k)$ ). The smoothing of $T(t)$ at
scale $r>0$ is defined as
$T_r(t)=\frac{1}{r}\int_{t-r/2}^{t+r/2}T(s)ds$. We consider the
processes $X_r(t)=\frac{T_r(t)}{T_1(t)}$, $t\in [0,1]$. The
empirical multifractal function $K(q)$ can be defined as the power
exponents if the following expectation behaves like \cite{Anh01}
\begin{equation}
E(X_r^q(t))\ \propto\ r^{K(q)}.
 \label{eq1}
\end{equation}
If we consider smoothing at discrete scales $r_j$, $j=1,2,\cdots$,
then from Eq. (1), the empirical $K(q)$ function (denoted as
$K_d(q)$) for the data can be obtained by
\begin{equation}
K_d(q)=\lim_{j \rightarrow \infty} \frac{\ln E(X_{r_j}^q)}{-\ln
r_j}.
 \label{eq2}
\end{equation}
Hence the empirical $K(q)$ function $K_d(q)$ can be estimated from
the slopes of $E(X_r^q)$  against the scale ratio 1/r in a log-log
plane. In this paper, we adopt Eq. (2) to obtain $K_d(q)$ of our
rainfall data.  If the curve $K_d(q)$ versus $q$ is a straight
line, the data set is monofractal. However, if this curve is
convex, the data set is multifractal \cite{Halsey86}.

The universal multifractal model (UMM) proposed by Schertzer and
Lovejoy \cite{Schertzer87} assumes that the generator of
multifractals was a random variable with an exponentiated extremal
L\'{e}vy distribution. Thus, the theoretical scaling exponent
function $K(q)$ for the moments $q\geq 0$ of a cascade process is
obtained according to \cite{Schertzer87, Tessier93,
Garcia-Marin08, Serinaldi10}:
\begin{equation}
K(q)=qH+\left\{
\begin{array}{ll}
C_{1}(q^{\alpha }-q)/(\alpha -1), & \alpha \neq 1, \\
C_{1}q\log (q), & \alpha =1,
\end{array}%
\right.
\end{equation}%
in which the most significant parameter $\alpha \in \lbrack 0,2]$ is the L%
\'{e}vy index, which indicates the degree of multifractality (i.e.
the deviation from monofractality). $C_1\in [0,d]$, with $d$ being
the dimension of the support ($d =1$ in our case), describes the
sparseness or inhomogeneity of the mean of the process
\cite{Garcia-Marin08}. The parameter $H$ is called the
non-conservation parameter since $H\neq 0$ implies that the
ensemble average statistics depend on the scale, while $H=0$ is a
quantitative statement of ensemble average conservation across the
scales (e.g., \cite{Serinaldi10}).

Although the double trace moment (DTM) technique \cite{Schmitt92,
Lavallee93} has been widely used to estimate the parameters $H$,
$C_{1}$ and $\alpha $ in geophysical research, it is complicated
and the goodness of fit of the empirical $K(q)$ functions depends
on that of exponent $\beta$ of the power spectrum, and sometimes
the fitting of $K(q)$ is not satisfactory (e.g., \cite{Olsson96,
Garcia-Marin08, Serinaldi10}). In this paper, we adopt a method in
\cite{Yu12} and is similar to that proposed in \cite{Anh01}. If we
denote $K_{T}(q)$ the $K(q)$ function defined by Eq. (3), we
estimate the parameters by solving the least-squares optimization
problem \cite{Yu12}
\begin{equation}
\min_{H,C_{1},\alpha
}\sum_{j=1}^{J}[K_{T}(q_{j})-K_{d}(q_{j})]^{2}. \label{eq6}
\end{equation}%
In our analysis, we take $q_j=j/3$ for $j=1,2,...,30$.

\subsection{Multifractal detrended fluctuation analysis}

 We outline the MF-DFA
procedure used here according to the procedure described in
\cite{Kantelhardt02}.

Suppose that $x_k$ is a series of length $N$.  First we determine
the 'profile' $Y(i)=\sum_{k=1}^i [x_k-\langle x\rangle],\ i = 1,
\cdots, N$, where $\langle x\rangle$ is the mean of $\{x_k\}$. For
an integer $s>0$, we divide the profile $Y(i)$ into $N_s=int(N/s)$
non-overlapping segments of equal lengths $s$, where $int(N/s)$ is
the integer part of $N/s$. Since the length $N$ of the series is
often not a multiple of the timescale $s$ under consideration,
there may remain a slack at the end of the profile. In order not
to disregard this short part of the series, the same procedure is
repeated starting from the opposite end. Thus, $2N_s$ segments are
obtained altogether. Now we can calculate the local trend for each
of the $2N_s$ segments by a least squares linear fit of the
series, then determine the variance $ F^{2}(s,\nu)$ for $
\nu=1,\cdots,2N_s$ \cite{Kantelhardt02}. Then the $q$th-order
fluctuation function is defined as $F_{q}(s)=\left[
\frac{1}{2N_s}\sum_{\nu=1}^{2N_s}( F^{2}(s,\nu))
^{q/2}\right]^{1/q}$, where $q\neq 0$. Finally we determine the
scaling behavior
\begin{equation}
F_{q}(s)\ \propto \ s^{h\left( q\right) }.  \label{eq7}
\end{equation}%
of the fluctuation functions by analyzing the log-log plot of
$F_q(s)$ versus $s$ for each value of $q$. The exponent $h(q)$  is
commonly referred to as the generalized Hurst exponent. The MF-DFA
is suitable for both stationary and nonstationary time series
\cite{Kantelhardt02}. We denote $\tilde{H}$ the Hurst exponent of
time series. The range $0.5 < \overline{H} < 1$ indicates long
memory or persistence; and the range $0 < \overline{H} < 0.5$
indicates short memory or anti-persistence. For uncorrelated
series, the scaling exponent $\overline{H}$ is equal to 0.5.
Assuming the setting of fractional Brownian motion, Movahed {\it
et al}. \cite{Movahed06} proved the relation $\overline{H}
=h(2)-1$ between $\overline{H}$ and the exponent $h(2)$ for small
scales. In the case of fractional Gaussian noise, it was shown
that $h(2)=\overline{H}$ \cite{Movahed06}. Hence we can use the
value of $\overline{H}$ calculated from $h(2)$ to detect the
nature of memory in time series under the assumption of fractional
Gaussian noise or fractional Brownian motion.

In the case of a power law, the power spectrum $S(f)$ is related
to the frequency $f$ by $S(f)\propto (1/f)^{\beta}$. The exponents
$h(2)$ and $\beta$ are related to each other by the equation
$h(2)=(1+\beta)/2$ \cite{Matsoukas00, Havlin88}. As pointed out by
Lovejoy {\it et al}. \cite{Lovejoy08}, the relationship between
 mass exponent $\tau(q)$, which is based upon the standard partition function
multifractal formalism \cite{Halsey86}, and $K(q)$ is
\begin{equation}
\tau(q)=(q-1)-K(q), \label{eq6}
\end{equation}
for 1-dimensional data. For a conservative process,
Koscielny-Bunde {\it et al}. \cite{Koscielny-Bunde06} pointed out
the relationship between $h(q)$ and $K(q)$ as
\begin{equation}
qh(q)=qh(1)-K(q). \label{eq7}
\end{equation}%
By combining Eqs. (6) and (7), we get \cite{Zhou11}
\begin{equation}
\tau(q)=q(h(q)-h(1))+q-1. \label{eq8}
\end{equation}%

\section{Results and discussion}

In this study, we apply the above methods to examine the
multifractal properties of daily rainfall data in Pearl River
basin over time as a regional case study. At each of the 41
stations in Pearl River basin, daily rainfall data over the period
from 1 January 1960 to 31 December 2005 consist of 16,802
observations. The information on location and elevation of the 41
stations in Pearl River basin is given in Table 1 (we list the
stations according to the deceasing order of their elevations).
According to the elevation, we can divide the stations into three
groups (Group 1 with elevation higher than 1000m, Group 2 with
elevation between 200m to 1000m, Group 3 with elevation lower than
200m). The daily rainfall data of Station 56691 and Station 57922
(in the Pearl River basin) over the entire study period are shown
in Figure 2 as examples.

First, we computed the empirical $K(q)$ curves of all daily
rainfall data via Eq.(2) by taking values for $r_j$  from 0.0010
to 0.056 (corresponding to time scale from 180-960 days) for data
in Pearl River basin because the power-law relation in Eq.(2) in
these time scale ranges becomes linear. An example for obtaining
the empirical $K(q)$ curves is given in Figure 3. The empirical
$K(q)$ curves of the rainfall data in two stations are shown in
Figure 4 (the dotted lines) as examples. We observed that all the
empirical $K(q)$ curves of the rainfall data in all stations are
not straight lines (i.e. are convex lines) like those in Figure 4.
This suggests that all daily rainfall time series have
multifractal behavior in the time scale range from 180 to 960
days. In order to use the UMM (i.e. Eq. (3)) to fit the empirical
$K(q)$ curves, we use the function {\it fminsearch} in MATLAB to
solve the optimization problem (Eq.(4)) and obtain the estimates
of parameters $H$, $\alpha$  and $C1$ (we set 0.5, 0.5, 0.5 as the
initial values of these three parameters, respectively). The
estimated values of parameter $\alpha$ for stations in the Pearl
River basin are given in Table 1. We found that the theoretical
$K(q)$ curves based on the UMM fit exceedingly well the empirical
$K(q)$ curves of the rainfall data in all stations. We plot two
fitted theoretical $K(q)$ curves in Figure 4 (the continuous
lines) as illustrations. From the estimated values of $H$, $C_1$
and $\alpha$, we find that $H \in [-0.0459, 0.0196]$ with mean
value $-0.0085 \pm 0.0126$, $C_1 \in [0.0867, 0.2665]$ with mean
value $0.1631\pm 0.0385$, and $\alpha \in [0.6213, 1.6072]$ with
mean value $1.0236 \pm 0.2141$ for stations in Pearl River basin.
The values of $H$ with mean value $-0.0085\pm 0.0126$ for these
daily rainfall data are close to zero, indicating that they
correspond to conserved fields which is consistent with previously
published results (e.g., [26-29]). Since the values of $\alpha$
are fairly large (far from the monofractal value of zero), it
again confirms that all daily rainfall time series in Pearl River
basin have multifractal behavior in the time scale range from 180
to 960 days. The values of $C_1$ with mean value $0.1631\pm
0.0385$ indicate that the conserved multifractal daily rainfall is
not too sparse \cite{Tessier93}, which can be compared with
previously published results \cite{Olsson96, Lima99}.

Second, we employed the MF-DFA to analyze the rainfall data. There
are usually seasonal variations in rainfall data. In order to get
the long term correlations correctly, the data need to be
deseasonalized before we can perform the MF-DFA [39, 40, 56-58].
In this paper, the deseasonalized rainfall $z_i$  ($i =
1,2,\cdots,N$, $N$ is the total number of data points) are
obtained by subtracting the mean daily rainfall $\overline{x_i}$
from the original rainfall $x_i$ and normalized by variance at
each calendar date [40, 56-58], i.e.,
\begin{equation}
z_i=(x_i-\overline{x_i})/(\overline{x_i^2}-\overline{x_i}^2).
\label{eq9}
\end{equation}
The deseasonalized rainfall was analyzed with MF-DFA. Here we
calculated $h(q)$ over the scale range of 10 to 87 days for all
values of $q$ because the log-log plot of $F_q(s)$  versus $s$ for
each value of $q$ in this time scale range becomes linear. An
example for obtaining the empirical $h(q)$ curve is given in
Figure 5. The empirical $h(q)$ curves of the rainfall data in two
stations are shown in Figure 6 as examples. We observed that all
the empirical $h(q)$ curves of the rainfall data in all stations
we considered are not straight lines (i.e. are convex lines) like
those in Figure 6. This suggests that all daily rainfall time
series have multifractal behavior in the time scale range from 10
to 87 days. Usually the value of $\Delta h(q)$ (defined as $\max
\{h(q)\}-\max \{h(q)\}$) is used to characterize the
multifractality of time series. The estimated values of $h(1)$,
$h(2)$ and $\Delta h(q)$ for stations in Pearl River basin are
given in Table 1. From Table 1, we find that $h(2)\in [0.5248,
0.6436]$ with mean value $0.5891\pm 0.0275$, $\Delta h(q) \in
[0.3724, 0.8851]$ with mean value $0.5681\pm 0.1210$ for stations
in Pearl River basin. The values of $\Delta h(q) \in [0.3724,
0.8851]$ obtained by us with mean value $0.5681\pm 0.1210$ (far
from the monofractal value of zero) for stations in Pearl River
basin again confirms that all daily rainfall time series in Pearl
River basin have multifractal behavior in the time scale range
from 10 to 87 days. It was reported that the scaling exponents of
rainfall obtained by DFA for the intermediate time scales (10.0 to
100.0-300.0 days) range in values from 0.62 to 0.89
\cite{Matsoukas00} without removing the seasonal trend in the
data. Later on, after removing the seasonal trend in the rainfall
data, Kantelhardt {\it et al.} \cite{Kantelhardt06} found that
most precipitation records exhibit no long-term correlations
($h(2)\approx 0.55$), the mean value is $h(2)=0.53 \pm 0.04$. The
values of $h(2)\in [0.5248, 0.6436]$ obtained by us with mean
value $0.5891\pm 0.0275$ for stations in Pearl River basin
consists with the result that precipitations are mainly
uncorrelated reported in \cite{Kantelhardt06}.

   It is also interesting to test the relationship between $K(2)$ and
$h(2)$ given by Eq. (7), i.e. whether $K(2)=2[h(1)-h(2)]$ holds.
We denote $K'(2)$ to be $2[h(1)-h(2)]$. The estimated values of
$K'(2)$ for stations in Pearl River basin are given in Table 1.
From Table 1, we find that $K'(2)\in [0.1960, 0.5300]$ with mean
value $0.2980\pm 0.0728$ for stations in Pearl River basin. We
find from Table 1 that the values of $K'(2)$ are quite different
from those of $K(2)$, this because that they are estimated for
different time scale ranges.

Last, we want to see whether the parameters from these MFAs of
daily rainfall can reflect some spatial or geographical
characteristics of the stations in Pearl River basin. In other
words, we would like to explore the spatial dimension of rainfall
variability in the basin. In particular, we are interested in
finding out whether rainfall variations over time are related to,
for example, the topography of the basin. A scrutiny of the
parameters $H$, $\alpha$ and $C_1$ in UMM, $K(2)$ in the $K(q)$
curves, and $h(2)$ from MF-DFA show that there exhibit some
correlations between rainfall regime and basin characteristics
such as topography. In fact, we found that the parameter $K(2)$,
which is related to the correlation dimension $D(2)$ via
$D(2)=1-K(2)$, of the daily rainfall data reflects some spatial
and geographical features of the stations in the basin. First,
K(2) and elevation series are negatively correlated. The value of
the correlation coefficient between $K(2)$ and elevation is up to
-0.4995 in the Pearl River basin as shown in Figure 7. The
possible trend is that the higher the elevation at which a station
is located, the smaller the value of $K(2)$ becomes and the closer
it is to 0.0 (so also the larger the value of $D(2)$ becomes and
the closer it is to 1.0). According to the elevation, we can
divide the stations into three groups (Group 1 with elevation
higher than 1000m, Group 2 with elevation between 200m to 1000m,
Group 3 with elevation lower than 200m). We found that $K(2)$ of
Group 1 have mean value $0.1927\pm 0.0110$, that of Group 2 have
mean value $0.2000\pm 0.0181$ and that of Group 3 have mean value
$0.2155\pm 0.0202$. One can see that the mean value of $K(2)$ of
these three groups become larger with decreasing of elevation. We
also notice that rainfall stations at higher elevations in the
northwestern side of the basin similarly tend to have smaller
$K(2)$ values in comparison with stations at lower elevations in
the southeastern side. Using the wavelet analysis on the monthly
precipitation data in Pearl River basin, Niu \cite{Niu13} recently
found that, apart from the high variability for the less than
1-year period, the high wavelet power in the dominant band
(0.84-4.8 years) for the first and second modes (especially for
northwest part and east part of Pearl River basin) reflects
long-term precipitation variability. Niu \cite{Niu13} explained
that the northwest region has the highest altitudes, and therefore
it is influenced by the topographic rain shadow  with respect to
the prevailing storm tracks; while the east region is close to the
South China Sea which is subjected to convective movement of water
by semitropical hurricanes and typhoons.

\section{Conclusion}

Multifractal analysis is a useful method to characterize the
heterogeneity of both theoretical and experimental fractal
patterns. As a regional case study, numerical results obtained
from the universal multifractal approach and MF-DFA on the daily
rainfall data in Pearl River basin show that these time series
have multifractal behavior in two different time scale ranges. It
is found that the empirical $K(q)$ curves of the daily rainfall
time series can be fitted very well by the UMM. The estimated
values of $H$ for these daily rainfall data are close to zero,
indicating a correspondence to the conserved fields.

After removing the seasonal trend in the rainfall data, the
estimated values of $h(2)$ indicate that the daily rainfall time
series in Pearl River basin exhibit no long-term correlations.

It is found that $K(2)$ and elevation series are negatively
correlated. It shows a relationship between topography and
rainfall variability.

\section*{Acknowledgements}

This project was supported by Geographical Modelling and
Geocomputation Program under the Focused Investment Scheme of The
Chinese University of Hong Kong, and the Earmarked grant
CUHK405308 of the Research Grants Council of the Hong Kong Special
Administrative Region; the Natural Science Foundation of China
(Grant no. 11071282 and 11371016), the Chinese Program for
Changjiang Scholars and Innovative Research Team in University
(PCSIRT) (Grant No. IRT1179), the Research Foundation of Education
Commission of Hunan Province of China (grant no. 11A122), the
Lotus Scholars Program of Hunan province of China.

\pagebreak
\begin{table}[tbp]
\caption{ \footnotesize The geographical information of the
rainfall stations and estimated multifractal parameters of the
daily rainfall data in Pearl River basin. We list the stations
according to the deceasing order of their elevations.}
\label{ZJ_para}
\begin{center}
{\scriptsize
\begin{tabular}{|c|c|c|c|c|c|c|c|c|c|c|}
\hline  & Station & Long. & Lat. & Elev. &   &   &   &   &  &\\
 Group & name & ( $^{\circ}$) & ( $^{\circ}$) & (m) &
$\alpha$ &  $h(1)$ & $\Delta h(q)$ &$h(2)$ & $K(2)$ & $K'(2)$ \\
\hline
  & 56691 & 104.28 & 26.87 & 2237.5 &  0.9905 & 0.7239 & 0.4602 & 0.6106  & 0.1780 & 0.2266\\ \cline{2-11}
  & 56786 &  103.83 & 25.58 &  1998.7 &  1.2019 & 0.7279 & 0.6758 & 0.5666 & 0.1970  & 0.3226\\ \cline{2-11}
  & 56875 &  102.55 & 24.33 &  1716.9 &  0.9807 & 0.7898 & 0.8851 & 0.5248 & 0.1951 & 0.5300\\ \cline{2-11}
Group 1 & 56886 &  103.77 & 24.53 &  1704.3 &  0.9563 & 0.7560 & 0.7340 & 0.5615 & 0.1877 & 0.3890\\
\cline{2-11}
(with Elev. &  57806 &  105.90 & 26.25 &  1431.1 &  1.1437 & 0.6958 & 0.3929 & 0.5978 & 0.1886 & 0.1960\\
\cline{2-11}
$\ge 1000$m  & 57902 &  105.18 & 25.43 &  1378.5 &  1.0583 & 0.6908 & 0.4169 & 0.5797 & 0.1874 & 0.2222\\
\cline{2-11}
  & 56985 &  103.38 & 23.38 &  1300.7 &  0.9451 & 0.7567 & 0.4517 & 0.5805 & 0.2159 & 0.3524\\ \cline{2-11}
  & 57922 &  107.55 & 25.83 &  1013.3 &  1.1786 & 0.6932 & 0.3905  & 0.5902 & 0.1919 & 0.2060\\
\hline
 & 59209 &  105.83 & 23.42 &  794.10 &  1.0300 & 0.7271 & 0.5094 & 0.5717 & 0.1745 & 0.3108\\ \cline{2-11}
 & 59218 &  106.42 & 23.13 &  739.90 &  1.1541 & 0.7188 & 0.6016 & 0.5779 & 0.1697 & 0.2818\\ \cline{2-11}
 Group 2 & 57906 &  106.08 & 25.18 &  566.80 &  0.9873 & 0.7125 & 0.4786 & 0.5816 & 0.2072 & 0.2618\\ \cline{2-11}
 (with Elev. & 59021 &  107.03 & 24.55 &  484.60 & 0.8460 & 0.6927 & 0.5752 & 0.5515 & 0.2009 & 0.2824\\
\cline{2-11}
between  & 57916 &  106.77 & 25.43 &  440.30 &  0.9970 & 0.7324 & 0.6086 & 0.5714 & 0.2105 & 0.3220\\
\cline{2-11}
 200m to  & 59102 &  115.65 & 24.95 &  303.90 &  1.0614 & 0.7619 & 0.6160 & 0.6034 & 0.2184 & 0.3170\\
\cline{2-11}
1000m) & 57932 &  108.53 & 25.97 &  285.70 &  1.0908 & 0.7102 & 0.5882 & 0.5807 & 0.2045 & 0.2590\\
\cline{2-11}
 & 59096 &  114.48 & 24.37 &  214.80 &  1.0230 &  0.7622 & 0.5354 & 0.6279 & 0.2145 & 0.2686\\
\hline
 & 59211 &  106.60 & 23.90 &  173.50 &  0.7283 & 0.7395 & 0.5026 & 0.5893 & 0.2222 & 0.3004\\ \cline{2-11}
 & 59037 &  108.10 & 23.93 &  170.80 &  1.2448 & 0.7338 & 0.5106 & 0.6168 & 0.2316 & 0.2340\\ \cline{2-11}
 & 57957 &  110.30 & 25.32 &  164.40 &  0.8394 & 0.7308 & 0.7217 & 0.5804 & 0.2192 & 0.3008\\ \cline{2-11}
 & 59058 &  110.52 & 24.20 &  145.70 &  1.0215 & 0.7190 & 0.3825 & 0.6176 & 0.1916 & 0.2028\\ \cline{2-11}
 & 57996 &  114.32 & 25.13 &  133.80 &  0.8723 & 0.7465 & 0.5241 & 0.6200 & 0.1991 & 0.2530\\ \cline{2-11}
 & 59417 &  106.85 & 22.33 &  128.80 &  1.4062 & 0.7411 & 0.5340 & 0.6131 & 0.1880 & 0.2560\\ \cline{2-11}
 & 59431 &  108.22 & 22.63 &  121.60 &  1.1208 & 0.7404 & 0.4751 & 0.6208 & 0.2169 & 0.2392\\ \cline{2-11}
 & 57947 &  109.40 & 25.22 &  121.30 &  0.8441 & 0.7265 & 0.5512 & 0.5933 & 0.2393 & 0.2664 \\ \cline{2-11}
Group 3 & 59265 &  111.30 & 23.48 &  114.80 &  1.4781 & 0.7321 & 0.5647 & 0.5843 & 0.2116 & 0.2956\\
\cline{2-11}
(with Elev. & 59065 &  111.53 & 24.42 &  108.80 &  0.8759 & 0.7425 & 0.4271 & 0.6272 & 0.1996 & 0.2306\\
\cline{2-11}
 $\le 200$m) & 59072 &  112.38 & 24.78 &  98.30  &  1.0028 & 0.7403 & 0.4047 & 0.6287 & 0.1949 & 0.2232\\
\cline{2-11}
 & 59046 &  109.40 & 24.35 &  96.80  &  0.8041 & 0.7196 & 0.6851 & 0.5592 & 0.2191 & 0.3208\\ \cline{2-11}
 & 59242 &  109.23 & 23.75 &  84.90  &  0.7554 & 0.7342 & 0.5614 & 0.6077 & 0.2200 & 0.2530\\ \cline{2-11}
 & 59087 &  113.53 & 23.87 &  68.60  &  1.1332 & 0.7595 & 0.6136 & 0.6126 & 0.2201 & 0.2938\\ \cline{2-11}
 & 59082 &  113.60 & 24.68 &  61.00  &  0.9246 & 0.7466 & 0.6232 & 0.6155 & 0.2060 & 0.2622\\ \cline{2-11}
 & 59271 &  112.43 & 23.63 &  57.30  &  1.2472 & 0.7295 & 0.5859 & 0.5820 & 0.1889 & 0.2950\\ \cline{2-11}
 & 59462 &  111.57 & 22.77 &  53.30 &   1.3551 & 0.7411 & 0.5056 & 0.5901 & 0.2097 &  0.3020\\ \cline{2-11}
 & 59254 &  110.08 & 23.40 &  42.50  &  0.7011 & 0.7496 & 0.3724 & 0.6436 & 0.1978 & 0.2120\\ \cline{2-11}
 & 59278 &  112.45 & 23.03 &  41.00  &  1.6072 & 0.7472 & 0.7785 & 0.5413 & 0.1945 & 0.4118 \\ \cline{2-11}
 & 59287 &  113.33 & 23.17 &  41.00  &  1.1680 & 0.7467 & 0.6313 & 0.5675 & 0.2113 & 0.3584\\ \cline{2-11}
 & 59293 &  114.68 & 23.73 &  40.60  &  0.9206 & 0.7711 & 0.5821 & 0.6040 & 0.2570 & 0.3342\\ \cline{2-11}
 & 59294 &  113.83 & 23.33 &  38.90  &  0.6213 & 0.7641 & 0.7561 & 0.5460 & 0.2104 & 0.4362\\ \cline{2-11}
 & 59478 &  112.78 & 22.25 &  32.70 &   0.8213 & 0.7829 & 0.6593 & 0.6031 & 0.2476 & 0.3596\\ \cline{2-11}
 & 59298 &  114.42 & 23.08 &  22.40  &  0.7185 & 0.7566 & 0.6993 & 0.5558 & 0.2308 & 0.4016\\ \cline{2-11}
 & 59493 &  114.10 & 22.55 &  18.20 &   1.1102 & 0.7695 & 0.7200 & 0.5552 & 0.2600 & 0.4286\\ \cline{2-11}
\hline mean & & & & & 1.0236 & 0.7381 & 0.5681 &0.5891 & 0.2080 &
0.2980 \\
$\pm$ std & & & & & $\pm$0.2141 & $\pm$0.0234 &
$\pm$0.1210 & $\pm$0.0275 & $\pm$0.0205 & $\pm$0.0728 \\
\hline
\end{tabular} }
\end{center}
\end{table}

\begin{figure}[bp]
\centerline{\epsfxsize=12cm\epsfbox{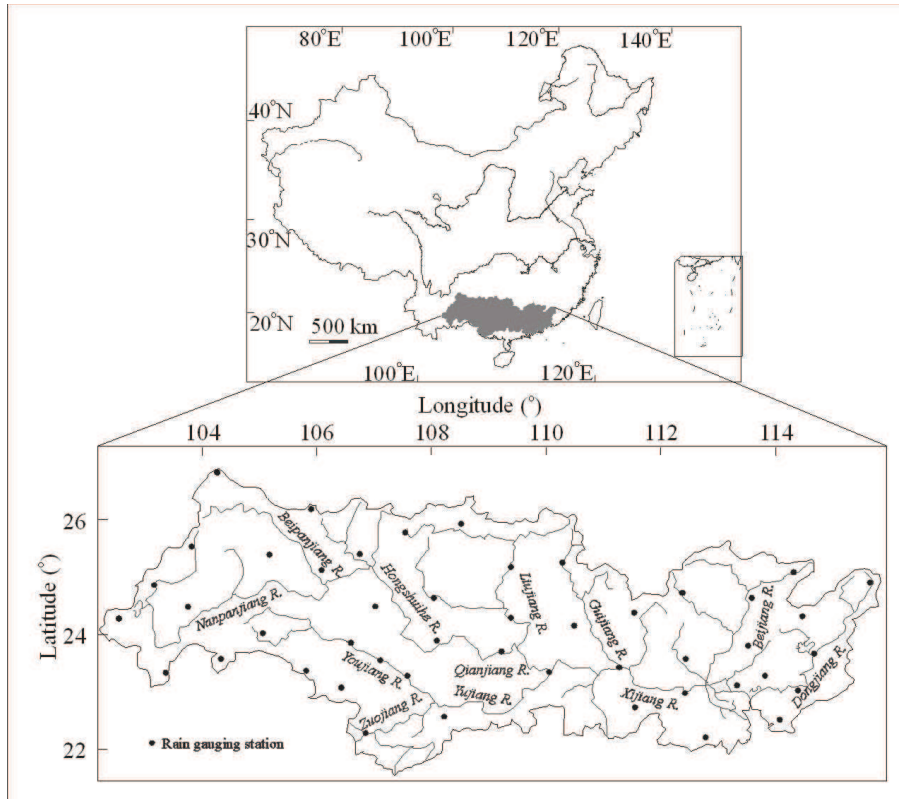}}
\caption{Location of the rain gauge stations in the Pearl River
basin, China. } \label{f1_map}
\end{figure}

\begin{figure}[bp]
\centerline{\epsfxsize=10cm\epsfbox{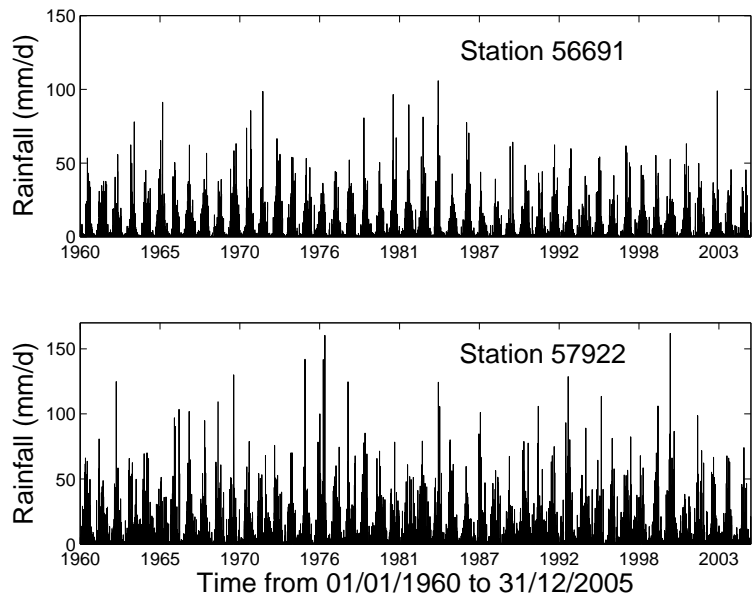}}
\caption{The daily rainfall data of station 56691 and Station
57922 in the Pearl River basin over the entire study period. }
\label{f2_data}
\end{figure}

\begin{figure}[bp]
\centerline{\epsfxsize=10cm\epsfbox{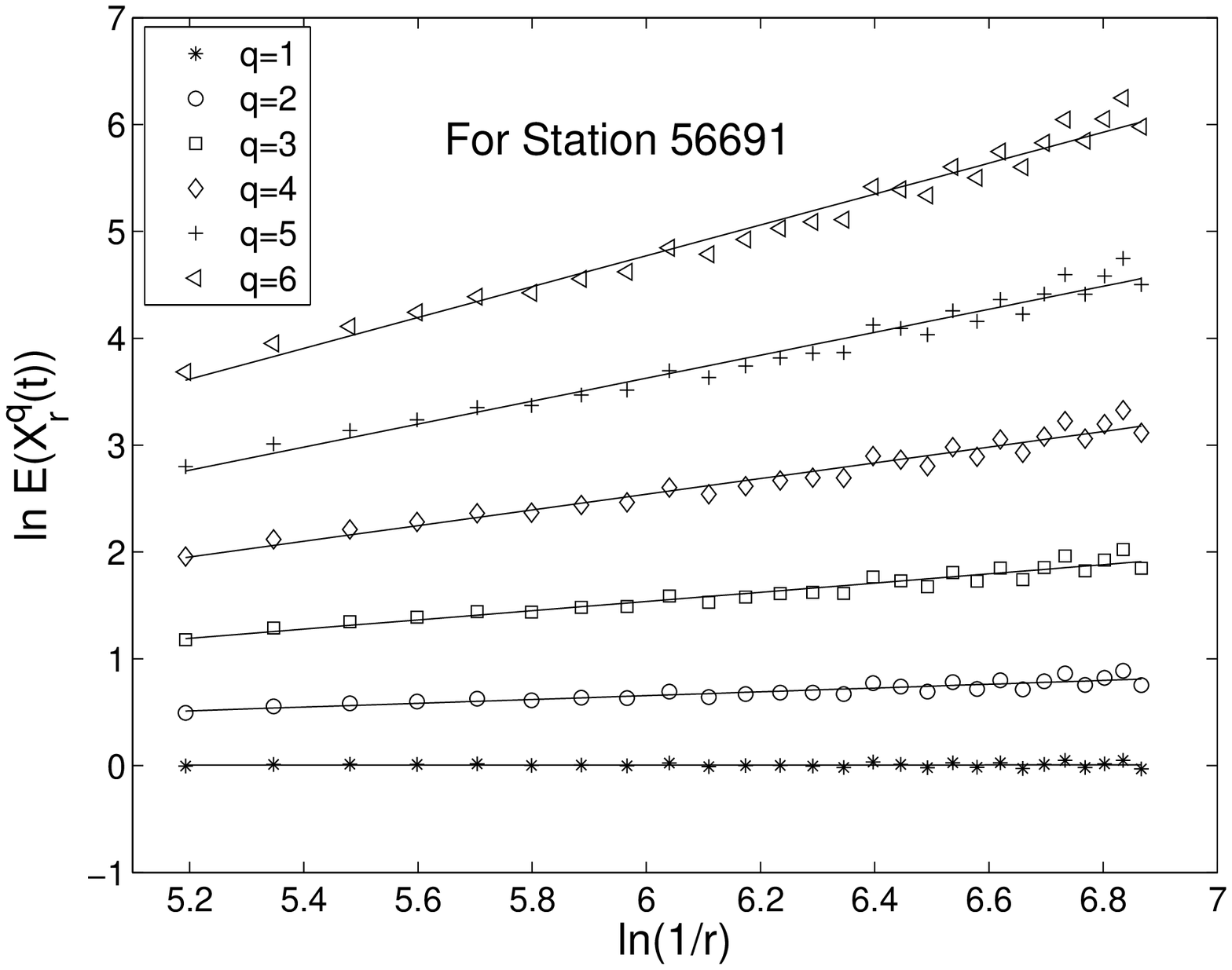}}
\caption{An example for obtaining the empirical $K(q)$ curve. }
\label{f3_Kq_slpoe}
\end{figure}

\begin{figure}[tbp]
\centerline{\epsfxsize=10cm\epsfbox{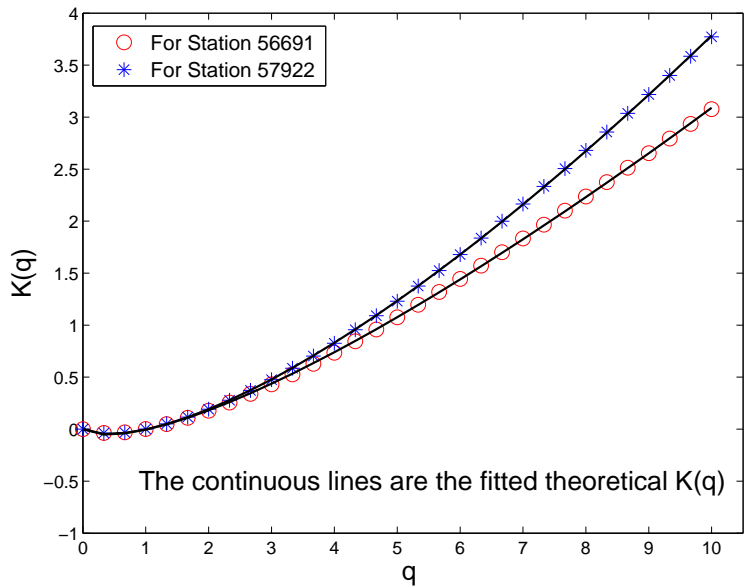}}
\caption{The $K(q)$ curves of daily rainfall data in two stations
(the dotted curves), and their fitted curves (continuous lines) by
the universal multifractal model. } \label{f4_Kq_fit}
\end{figure}

\begin{figure}[tbp]
\centerline{\epsfxsize=10cm\epsfbox{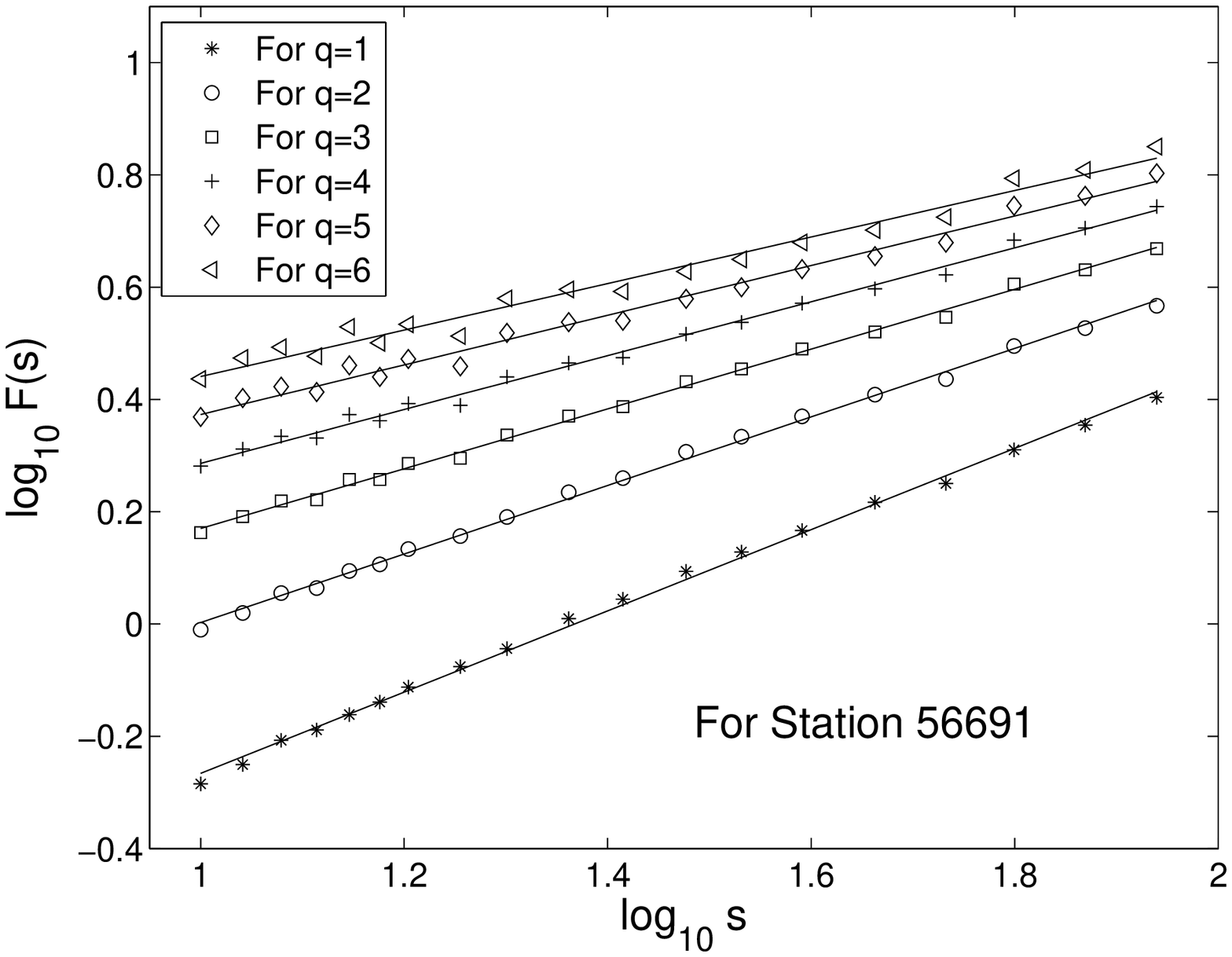}}
\caption{An example for obtaining the empirical $h(q)$ curve. }
\label{f5_hq_slope}
\end{figure}

\begin{figure}[tbp]
\centerline{\epsfxsize=10cm \epsfbox{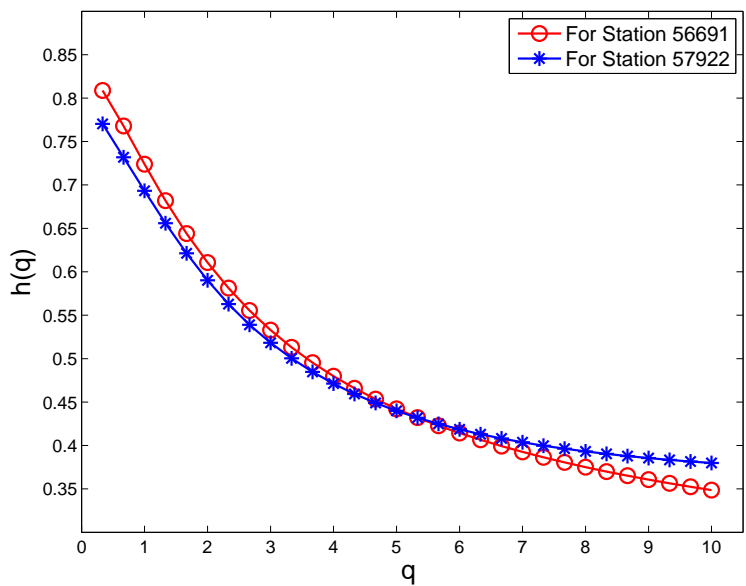}}
\caption{The $h(q)$ curves of daily rainfall data in two stations.
} \label{f6_hq_curves}
\end{figure}

\begin{figure}[tbp]
\centerline{\epsfxsize=10cm\epsfbox{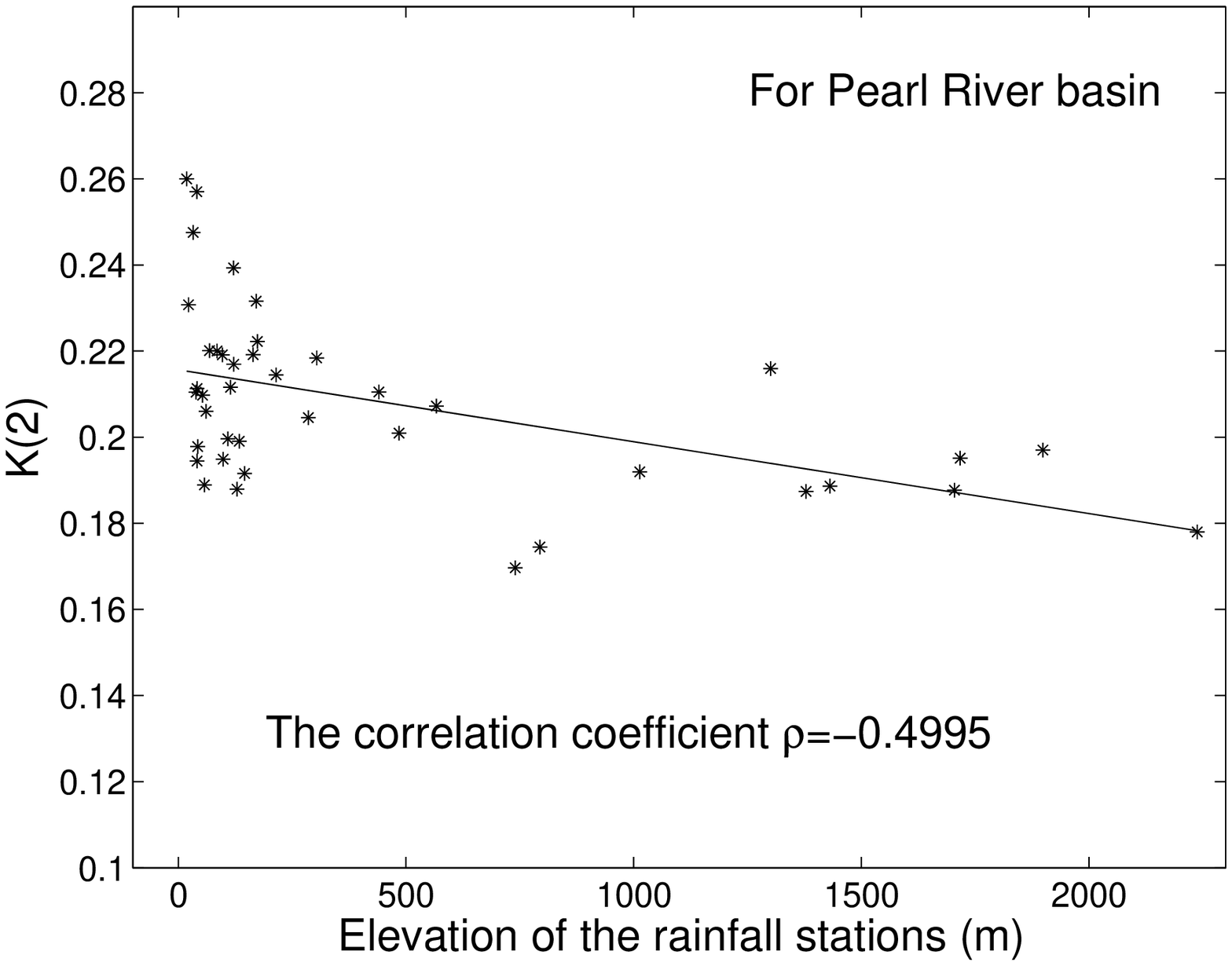}}
\caption{The correlation relationship between the elevation of the
rainfall stations and the $K(2)$ value of the rainfall time series
for the Pearl River basin. } \label{corr_coeff}
\end{figure}


\begin{thebibliography}{10}

{\footnotesize

\bibitem[1]{Walther02} G.R. Walther, E. Post, P. Convey, A. Menzel, C. Parmesank, T.J.C. Beebee, J.-M. Fromentin,
O. Hoegh-Guldberg and F. Bairlein, Ecological responses to recent
climate change. Nature 416 (2002) 389-395.

\bibitem[2]{Valencia10} J.L. Valencia, A.S. Requejo, J.M. Gasco, A.M. Tarquis, A universal multifractal description
applied to precipitation patterns of the Ebro River Basin, Spain.
Clim. Res. 44 (2010) 17-25.

\bibitem[3]{Zhang08}  Q. Zhang, C.-Y. Xu, Z. Zhang, Y.D. Chen, C.-L.
Liu, Spatial and temporal variability of precipitation maxima
during 1960-2005 in the Yangtze River basin and possible
association with large-scale circulation.
 J. Hydrol. 353 (2008) 215-227.

\bibitem[4]{Schertzer87} D. Schertzer and S. Lovejoy, Physical modeling and analysis of rain and clouds by anisotropic scaling of multiplicative processes.
J. Geophys. Res. 92(D8) (1987) 9693-9714.

\bibitem[5]{Gupta90} V.K. Gupta and E. Waymire, Multiscaling properties of spatial rainfall and river flow distributions.
J. Geophys. Res. 95(D3) (1990) 1999-2009.

\bibitem[6]{Over94} T.M. Over and V.K. Gupta, Statistical analysis of masoscale rainfall: dependence of a random cascade generator on large-scale forcing.
J. Appl. Meteorol. 33 (1994) 1526-1542.

\bibitem[7]{Schmitt98} F. Schmitt, S. Vannitsem and A. Barbosa, Modeling of rainfall time series using two-state renewal processes and multifractals.
J. Geophys. Res. 103(D18) (1998) 23181-23193.

\bibitem[8]{Svensson96} C. Svensson, J. Olsson and R. Berndtsson, Multifractal properties of daily rainfall in two different climates.
  Water Resour. Res. 32(8) (1996) 2463-2472.

\bibitem[9]{Sivakumar00} B. Sivakumar, Fractal analysis of rainfall observed in two different climatic regions.
Hydrol. Sci. J. 45(5) (2000) 727-738.

\bibitem[10]{Michele05} C. De Michele and P. Bernardara, Spectral analysis and modeling of space-time rainfall fields.
 Atmos. Res. 77 (2005) 124-136.

\bibitem[11]{Olsson93} J. Olsson, J. Niemczynowicz, and R. Berndtsson, Fractal analysis of high-resolution rainfall time series.
J. Geophys. Res. 98(D12) (1993) 23265-23274.

\bibitem[12]{Venugopal06} V. Venugopal, S.G Roux., E. Foufoula-Georgiou and A.
Arneodo, Revisiting multifractality of high-resolution temporal
rainfall using a wavelet-based formalism. Water Resour. Res. 42
(2006) W06D14.

\bibitem[13]{Molnar08} P. Molnar  and P. Burlando, Variability in the scale properties of high resolution precipitation data in the Alpine climate of Switzerland.
 Water Resour. Res. 44(10) (2008) W10404.

\bibitem[14]{Boni08}  G. Boni, A. Parodi and F. Siccardi, A new parsimonious methodology of mapping the spatial variability of annual maximum rainfall in mountainous environments.
  J. Hydrometeorol. 9(3) (2008) 492-506.

\bibitem[15]{Veneziano02} D. Veneziano and P. Furcolo, Multifractality of rainfall and scaling of intensity-duration-frequency curves.
Water Resour. Res. 38(12) (2002) 1306.

\bibitem[16]{Veneziano06} D. Veneziano, A. Langousis and P. Furcolo, Multifractality and rainfall extremes: A review.
 Water Resour. Res. 42 (2006) W06D15.

\bibitem[17]{Veneziano12} D. Veneziano  and C. Lepore, The scaling of temporal rainfall.
 Water Resour. Res. 48 (2012) W08516.

\bibitem[18]{Tessier93} Y. Tessier, S. Lovejoy and D. Schertzer, Universal multifractals: theory and observations for rain and clouds.
 J. Appl. Meteorol. 32(2) (1993) 223-250.

\bibitem[19]{Olsson96} J. Olsson and J. Niemczynowicz, Multifractal analysis of daily spatial rainfall distributions.
J. Hydrol. 187(1-2) (1996) 29-43.

\bibitem[20]{Perica96} S. Perica and E. Foufoula-Georgiou, Model for multiscale disaggregation of spatial rainfall based on coupling meteorological and scaling descriptions.
J. Geophys. Res. 101 (D21) (1996) 26347-26341.

\bibitem[21]{Menabde97}  M. Menabde, D. Harris, A. Seed, G. Austin and D. Stow,
Multiscaling properties of rainfall and bounded random cascades.
Water Resour. Res. 33(12) (1997) 2823-2830.

\bibitem[22]{Pandey98} G. Pandey, S. Lovejoy and D. Schertzer, Multifractal analysis of daily river
flows including extremes for basins of five to two million square
kilometres, one day to 75 years.
  J. Hydrol. 208(1-2) (1998) 62-81.

\bibitem[23]{Lima99} I. De Lima and J. Grasman, Multifractal analysis of 15-min and daily rainfall from a semi-arid region in Portugal.
 J Hydrol., 220 (1999) 1-11.

\bibitem[24]{Deidda99} R. Deidda, R. Benzi and F. Siccardi, Multifractal modeling of anomalous scaling laws in rainfall.
Water Resour. Res. 35 (6) (1999) 1853-1867.

\bibitem[25]{Lilley06} M. Lilley, S. Lovejoy, N. Desaulniers-Soucy, D. Schertzer,
Multifractal large number of drops limit in rain.  J. Hydrol. 328
(2006) 20-37.

\bibitem[26]{Lovejoy08} S. Lovejoy, D. Schertzer and V. Allaire, The remarkable wide range scaling of TRMM precipitation.
 Atmos. Res. 90 (2008) 10-32.

\bibitem[27]{Lovejoy12} S. Lovejoy, J. Pinel, D. Schertzer, The global space-time cascade structure of precipitation: Satellites,
gridded gauges and reanalyses.  Adv. Water Resour. 45 (2012)
37-50.

\bibitem[28]{Garcia-Marin08}  A.P. Garcia-Marin, F.J. Jimenez-Hornero, and J.L. Ayuso-Munoz,
Universal multifractal description of an hourly rainfall time
series from a location in southern Spain.  Atmosfera 21(4) (2008)
347-355.

\bibitem[29]{Serinaldi10} F. Serinaldi, Multifractality, imperfect scaling and hydrological properties of rainfall time series simulated
by continuous universal multifractal and discrete random cascade
models. Nonlin. Processes Geophys. 17 (2010) 697-714.

\bibitem[30]{Halsey86} T.C. Halsey, M.H. Jensen, L.P. Kadanoff, I. Procaccia and B.I.
Schraiman, Fractal measures and their singularities: the
characterization of strange sets.  Phys. Rev. A 33 (1986)
1141-1151.

\bibitem[31]{Kantelhardt02} J.W. Kantelhardt, S.A. Zschiegner, E. Koscielny-Bunde, S. Havlin,
A. Bunde and H.E. Stanley, Multifractal detrended fluctuation
analysis of nonstationary time series.  Physica A 316 (2002)
87-114.

\bibitem[32]{Peng92} C.K. Peng, S.V. Buldyrev, A.L. Goldberger, S. Havlin, F.
Sciortino, M. Simons, and H.E. Stanley, Long-range correlations in
nucleotide sequences.   Nature 356 (1992) 168-170.

\bibitem[33]{Peng94} C.K. Peng, S.V. Buldyrev, S. Havlin, M. Simons, H.E. Stanley and
A.L. Goldberger, Mosaic organization of DNA nucleotides.  Phys.
Rev. E 49 (1994) 1685-1689.

\bibitem[34]{Yu01} Z.G. Yu, V. Anh and B. Wang, Correlation property of length sequences based on
global structure of complete genome.  Phys. Rev. E 63 (2001)
011903.

\bibitem[35]{Yu06} Z.G. Yu, V.V. Anh, K.S. Lau and L.Q. Zhou,   Fractal and multifractal analysis of
hydrophobic free energies and solvent accessibilities in proteins.
Phys. Rev. E 73 (2006) 031920.

\bibitem[36]{Matsoukas00} C. Matsoukas, S. Islam, I. Rodriguez-Iturbe, Detrended fluctuation analysis of
rainfall and streamflow time series.  J. Geophys. Res. 105 (D23)
(2000) 29165-29172.

\bibitem[37]{Kantelhardt03} J.W. Kantelhardt, D. Rybski, S.A. Zschiegner, P. Braun, E.
Koscielny-Bunde, V. Livina, S. Havlin and A. Bunde,
Multifractality of river runoff and precipitation: comparison of
fluctuation analysis and wavelet methods.  Physica A 330 (1-2)
(2003) 240-245.

\bibitem[38]{Kantelhardt06} J.W. Kantelhardt, E. Koscielny-Bunde, D. Rybski, P. Braun, A.
Bunde and S. Havlin, Long-term persistence and multifractality of
precipitation and river runoff records.  J. Geophys. Res. 111(D1)
(2006) D01106.

\bibitem[39]{Koscielny-Bunde06} E. Koscielny-Bunde, J.W. Kantelhardt, P. Braun, A. Bunde and S.
Havlin, Longterm persistence and multifractality of river runoff
records: Detrended fluctuation studies.  J. Hydrol. 322(1-4)
(2006) 120-137.

\bibitem[40]{Li07} Z.W. Li and Y.K. Zhang, Quantifying fractal dynamics of groundwater systems with detrended fluctuation analysis,
J. Hydrol. 336(1-2) (2007) 139-146.

\bibitem[41]{Anh07} V. Anh, Z.G. Yu and J.A. Wanliss, Analysis of global geomagnetic variability.
 Nonlin. Processes Geophys. 14(6)
(2007) 701-708.

\bibitem[42]{Anh08} V.V. Anh, J.M. Yong and Z.G. Yu, Stochastic modeling of the auroral  electrojet index.
 J. Geophys. Res. 113 (2008)  A10215.

\bibitem[43]{Zhang09} Q. Zhang,
C.-Y. Xu, Z.G. Yu, C.-L. Liu, Y.D. Chen, Multifractal analysis of
streamflow records of the East River basin (Pearl River), China.
Physica A 388 (2009) 927-934.

\bibitem[44]{Zhang10} Q. Zhang, Z.G. Yu, C.-Y. Xu, V. Anh, Multifractal analysis of measure representation of flood/drought grade
series in the Yangtze Delta, China, during the past millennium and
their fractal model simulation.
 Int. J. Climatol. 30 (2010) 450-457.

 \bibitem[45]{Yu09} Z.G. Yu, V. Anh and R. Eastes, Multifractal analysis of geomagnetic storm and solar flare indices and their class dependence.
 J. Geophys. Res. 114 (2009) A05214.

\bibitem[46]{Yu10} Z.G. Yu, V. Anh, Y. Wang, D. Mao and J. Wanliss, Modeling and simulation of the horizontal component of the geomagnetic field by fractional
stochastic differential equations in conjunction with empirical
mode decomposition.  J. Geophys. Res. 115 (2010) A10219.

\bibitem[47]{Zhang2009a} Q. Zhang, C.Y. Xu, S. Becker, Z.X. Zhang, Y.Q. Chen, M. Coulibaly,
Trends and abrupt changes of precipitation maxima extremes in the
Pearl River basin, China. Atmos. Sci. Lett. 10 (2009) 132¨C144.

\bibitem[48]{Gemmer2011} M. Gemmer, T. Fischer, B. Su, L.L. Liu,  Trends of precipitation
extremes in the Zhujiang River Basin, South China. J. Clim. 24
(2011)750¨C761

\bibitem[49]{Anh01} V. Anh, K.S. Lau and Z.G. Yu, Multifractal characterisation of complete genomes.
 J. Phys. A: Math. Gen.
34(36) (2001) 7127-1739.

\bibitem[50]{Schmitt92} F. Schmitt, D. Lavallee, D. Schertzer and S.
Lovejoy, Empirical determination of universal multifractal
exponents in turbulent velocity fields. Phys. Rev. Lett., 68
(1992) 305-308.

\bibitem[51]{Lavallee93} D. Lavallee, S. Lovejoy, D. Schertzer and P. Ladoy,
Nonlinear variability and landscape topography: analysis and
simulation.  In: Fractals in Geography (N. Lam and L. De Cola,
Eds.) Prentice Hall, Englewood Cliffs, p158-192, 1993.

\bibitem[52]{Yu12} Z.G. Yu, V. Anh, R. Eastes and D.L. Wang, Multifractal analysis of solar flare indices and their horizontal visibility graphs.
  Nonlin. Processes Geophys.  19 (2012) 657-669.

\bibitem[53]{Movahed06} M.S. Movahed, G.R. Jafari, F. Ghasemi, S. Rahvar and M.R.R.
Tabar, Multifractal detrended fluctuation analysis of sunspot time
series.   J. Stat. Mech.: Theory Exper. 2 (2006) P02003.

\bibitem[54]{Havlin88} S. Havlin, R. Selinger, M. Schwartz,
H.E. Stanley, and A. Bunde, Random multiplicative processes and
transport in structures with correlated spatial disorder.
 Phys. Rev. Lett. 61(13) (1988) 1438-1441.

\bibitem[55]{Zhou11} Y. Zhou, Y. Leung and Z.G. Yu, Relationships of exponents in multifractal detrended fluctuation analysis and conventional multifractal analysis.
  Chin. Phys. B 20(9) (2011) 090507.

\bibitem[56]{Lawrence77} A.J. Lawrence, N.T. Kottegota, Stochastic modeling of riverflow time series.
  J. R. Stat. Soc., Ser. A (General) 140
(1) (1977) 1-47.

\bibitem[57]{Kocielny-Bunde96} E. Kocielny-Bunde, A. Bunde,  S. Havlin, Y. Goldreich,
Analysis of daily temperature fluctuations.  Physica A 231 (1996)
393-396.

\bibitem[58]{Livina03} V. Livina, Y. Ashkenazy, Z. Kizner, V. Strygin, A. Bunde, S.
Halvin, A stochastic model of river discharge fluctuations.
Physica A 330 (2003) 283-290.

\bibitem[59]{Niu13} J. Niu, Precipitation in the Pearl River
basin, South China: scaling, regional patters, and influence of
large-scale climate anomalies. Stoch. Environ. Res. Risk Assess.
27 (2013) 1253-1268.

}
\end{thebibliography}
\end{document}